# The growth of interest for astronomical X-ray polarimetry

**Frédéric Marin** [1]

[1] Université de Strasbourg, CNRS, Observatoire astronomique de Strasbourg, UMR 7550, F-67000 Strasbourg, France; frederic.marin@astro.unistra.fr

**Abstract:** Astronomical X-ray polarimetry was first explored in the end of the 60's by pioneering rocket instruments. The craze arising from the first discoveries on stellar and supernova remnant X-ray polarization led to the addition of X-ray polarimeters on-board of early satellites. Unfortunately, the inadequacy of the diffraction and scattering technologies required to measure polarization with respect to the constraints driven by X-ray mirrors and detectors, coupled to long integration times, slowed down the field for almost 40 years. Thanks to the development of new, highly sensitive, compact X-ray polarimeters in the beginning of the 2000's, the possibility to observe astronomical X-ray polarization is rising again and scientists are now ready to explore the high energy sky thanks to modern X-ray polarimeters. In the forthcoming years, several X-ray missions (both rockets, balloons and satellites) will open a new observational windows. A wind of renewal blows over the area of X-ray polarimetry and this paper presents for the first time a quantitative assessment, all based on scientific literature, of the growth of interest for astronomical X-ray polarimetry.

## 1. Introduction

Since the beginning of systematical studies of X-rays in 1895 [1] and the birth of observational X-ray astronomy in the early 60's [2], theoretical X-ray astronomy predicts a wealth of results from polarimetry but the conception of polarization-sensitive detectors took decades to be achieved. It was only in the end of the 60's that a couple of Lithium-block Thomson-scattering polarimeters was flown on sounding rockets, targeting the brightest X-ray source known at that time: Sco X-1 [3]. More advanced experiments were conducted on Aerobee-350 rockets, loaded with two instruments in one payload: a Lithium scattering polarimeter and a network of 4 Bragg crystal polarimeters. This experiment led to the first detection of X-ray polarization from the Crab nebula [4], with the systematic effects being mitigated by rotating the detectors. Further efforts were put on rockets and satellites, such as Intercosmos-1 which measured the X-ray polarization from solar flares [5] or Ariel 5, equipped with a Bragg crystal spectrometer-polarimeter that produced an upper limit of 7.7% at three-sigma confidence for Sco X-1 [6]. The OSO-8 mission led to the sole positive detection of non-solar X-ray polarization known so far, still in the Crab [7,8], with a linear polarization degree P = 19.2 ± 1.0 % and a polarization position angle θ = 156.4 ± 1.4° at 2.6 keV. A handful of upper limits were estimated for other objects but most of them are unfortunately of marginal significance. The field of X-ray polarimetry was opened and several missions were envisioned to pursue the first discoveries. The Stellar X-ray Polarimeter (SXRP) was planned to fly on the Russian Spectrum-X Gamma Mission in the early 1990s [9] but the collapse of the Soviet Union hampered this satellite to fly. More dramatically, HEAO-2/Einstein revolutionized the way X-ray observations were made: a cosmic source became a cluster of detected photons in an imaging detector in the focus of optics. Rotation was no more needed for spectroscopy or imaging and since then was considered as a costly complication. The technical contrast between spectroscopy/imaging and polarimetry (that needed rotation), and the sensitivity mismatching between the different techniques prevented the addition of complex, time-consuming, X-ray polarimeters in observatory missions. Even if rotation could be accomplished by rotating the detector on missions with telescopes (such as for SXRP), polarimeters were removed at different development stages (Einstein, AXAF) or from the very beginning (XMM, ATHENA) of many missions as the integration times required to measure polarized fluxes are much longer than for imaging or spectroscopy.

The modern breakthrough for X-ray polarization detectors was achieved in the beginning of the 2000's. The advent of electron-tracking polarimeters [10,11] allowed for compact detectors capable of imaging processes together with polarization measurement. In comparison with old diffraction or scattering polarimeters, the increase of sensitivity is without comparison: about a factor hundred is expected with respect to the polarimeters that flown on-board of OSO-8. Coupled to state-of-the-art X-ray optics, the field of X-ray polarimetry is now ready to open again after 40 years of silence. The forthcoming launch of space-born X-ray polarimeters [12] and the many balloon [13] and rocket [14] experiments that are currently planned clearly indicate a growing interest for X-ray polarimetry. This is precisely what this paper is intended for: to quantitatively show the enthusiasm of the research community for X-ray polarimetry. To do so, I will present in Sect. 2 the amount of registered publications on astronomical X-ray polarimetry, starting from the discovery of X-rays. I will then divide the database into subcategories to determine if all fields are growing at the same rate or if a specific domain is leading. Finally, I will conclude on the importance of X-ray polarimetry in Sect. 3.

## 2. Data mining and results

To identify and collect all individual papers on astronomical X-ray polarimetry, I used the SAO/NASA Astrophysics Data System (ADS). It is a digital library portal for researchers in Astronomy and Physics, operated by the Smithsonian Astrophysical Observatory (SAO) under a NASA grant. There are more than 13.4 million records covering publications in Astronomy, Astrophysics and general Physics. A search by keywords was mandatory in order to narrow down the number of relevant papers. The list of keywords used for data mining comprised the terms "X-ray", "Rontgen", "polarization", "polarized", "polarimeter", "polarimetry", "polarimetric" …, both in American English and British English. The resulting papers were examined individually to reject duplicates or false detections. The selected publications were stored in a database and classified per year and general subject. Six categories were created: "General", "Theory and instrumentation", "Solar, stellar and planets", "Strong gravity", "Strong magnetism" and "Galaxies (quiescent, clusters)". The "General" label stands for reviews, books and historical notes that are not developing new aspects or mathematical equations in the field. The "Theory and instrumentation" label includes papers on the nature of X-rays and their fundamental properties. It also includes mechanical and technical papers about the development of instruments that goes along the aforementioned theoretical papers. All publications, regardless of the journal rank (A, B …), were selected, together with lecture notes and proceedings. Only copies of seminar/conference presentations were rejected as they do not correspond to a publication *sensu stricto*. 838 papers (as of the beginning of January, 2018) were registered.

### 2.1. Evolution of publications on X-ray polarimetry

Figure 1 shows the results of the search and classification of papers relevant to the field of astronomical X-ray polarimetry. All subclasses of papers were merged, regardless of the scientific topic, in order to assess the growing interest of the community for high energy polarimetry. The resulting histogram covers the period January 1895 – January 2018. It immediately appears that the field stagnated for years between the discovery of X-rays and the launch of the first Lithium-block Thomson-scattering polarimeter, mounted inside an Aerobee-150 rocket. This experiment led to a first upper limit on the polarization of Sco X-1 [3] and can be considered as the onset of the First Age of astronomical X-ray polarimetry. The cornerstones which dot the years between 1895 and 1968 (first rocket-borne experiment performed in July 1968, in search of X-ray polarization from Sco X-1 between 5.0 and 16.8 keV) are the following: in "A" is the discovery of X-rays. In "B" is the first mention of X-ray polarization [15] and in "C" is a fundamental paper summarizing the full understanding of the nature of the gamma and X-rays [16]. In "D" is the flight of the aforementioned Aerobee-150 rocket and one can see the impact of this experiment onto the publication rate. The very next year, the publication number is found to be five times higher. Following celestial X-ray polarization measurement, the first measurements of polarization from non-thermal bremsstrahlung of solar flares is recorded in 1970 by the Intercosmos-1 mission (label "E") [5]. The number of publication kept increasing, up to 14 per year in 1972, which corresponds to the first non-solar astronomical X-ray polarization detection (label "F") [4]. As mentioned in the introduction, the target was the Crab Nebula and the polarimeters were mounted on an Aerobee-350 rocket. Shortly after, the 8th Orbiting Solar Observatory (OSO-8) was launched on June 1975. While OSO-8's primary objective was to observe the Sun, four instruments were dedicated to observations of other celestial X-ray sources brighter than a few milliCrab. In particular, the Graphite Crystal X-ray Spectrometer on-board allowed for X-ray polarization measurements in the 2-8 keV band (Field of View 3°). The analysis of 15 orbits of quick-look data on the Crab Nebula showed that the polarization and position angles at 2.6 and 5.2 keV were $15.7 \pm 1.5\%$ at $161.1 \pm 2.8°$ and $18.3 \pm 4.2\%$ at $155.5 \pm 6.6°$, respectively [7]. This is point "G" on Figure 1 and it corresponds to the golden era of the First Age of X-ray polarimetry. The number of publication on the subject started to decrease and reached a plateau of ~7 publications per year in the period 1980 – 2000. During this extent of time a number of significant but unfortunately unnoticed steps were conducted. In particular, highlighted by point "H", is the unfinished development of the Spectrum-X-Gamma mission and its fully developed SXRP.

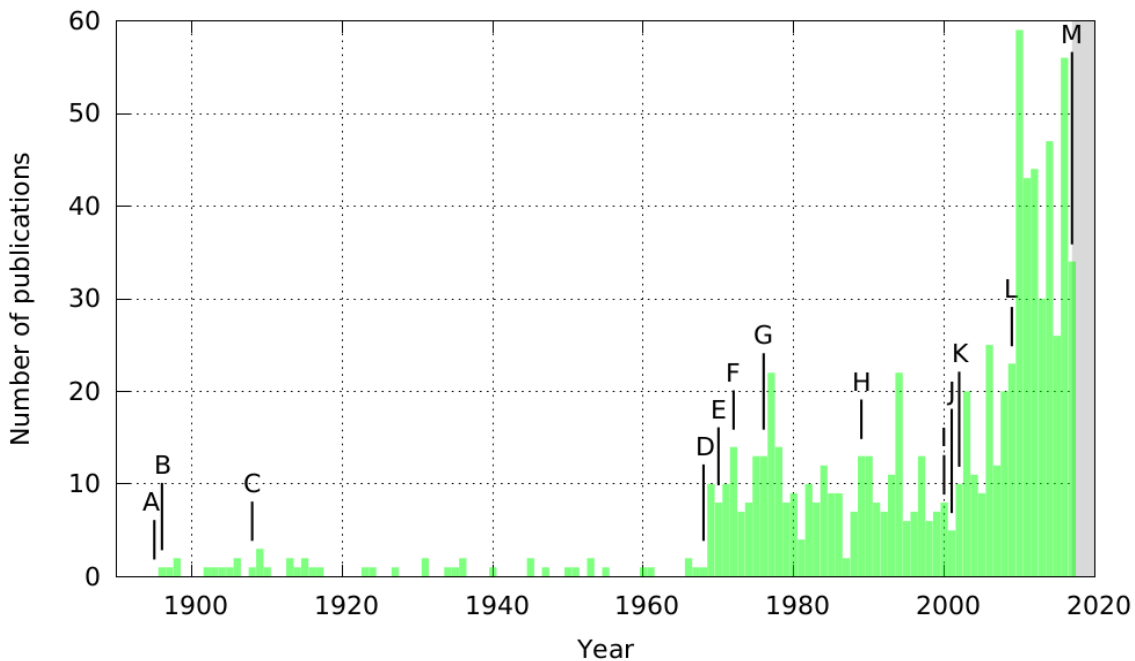

**Figure 1.** Number of publications focusing on astronomical X-ray polarimetry as a function of time. All types (refereed papers, proceedings, notes ...) and all scientific fields (solar physics, compact objects, ...) are merged. Letters indicate cornerstones in the history of X-ray polarimetry and are detailed in the text.

When the design of the SXRP was in an advanced status, a modulation in the photocurrent of a Caesium Iodide, Gold, and Aluminum photocathodes inpinged at grazing angles by 60-200 eV X-rays was found [17]. This modulation was much higher than what predicted by usual transfer programs. Even though this signal was later found to be an artifact, it was the starter point for a node of scientists and engineers, who begun to reconsider photoelectric polarimetry as a potential solution to increase the sensitivity of X-ray polarimeters. Hence, the beginning of the new millenium saw several major innovations and attempts in measuring X-ray polarimetry. In point "I" are X-ray polarimetry experiments with balloon borne gas proportional counters [18]. Detection of the X-ray polarization was successful only in a few sources, including Crab nebulae at 2.6keV and 5.2keV. Point "J" shows the development of the Gas Pixel Detector, achieved in 2001 [10] and that will carry out his first space flight in 2021. Point "K" stands for both the development of active-matrix pixel proportional counters [11] and the measurement of hard X-ray polarization of solar flares with RHESSI [19]. This is the beginning of a Second Age for X-ray polarimetry. From this point (circa 2005), the amount of publications started to rise again at a rapid pace. From 9 papers in 2005, the publication rate increased up to 59 in 2010. Since then, the averaged publication rate between 2010 and 2018 is ~ 42 papers per year. This important growth is the results of the development of modern X-ray polarimeters and the beginning of recognition by the NASA, ESA, CNSA, JAXA and other national space agencies that X-ray polarimetry represents a step forwards in the study of the high energy sky. In particular, the selection of the Gravity and Extreme Magnetism SMEX (GEMS) in 2009 was an important cornerstone (point "L" on Figure 1), but the mission was discontinued in May 2012. Point "M" is where we stand now: the Imaging X-ray Polarimetry Explorer (IXPE) mission was selected [12] and will fly in 2021; several balloon and rocket experiments are envisioned, and numerous other small or medium-sized spatial missions such as the enhanced X-ray Timing and Polarimetry mission (eXTP) are also planned [20].

*2.2. Publication per field*

We saw that the amount of publications is growing at a significant rate. More importantly, half of the publication in astronomical X-ray polarimetry were written after 2005. With such a high infatuation, it is interesting to check whether all sub-categories are following the same trend.

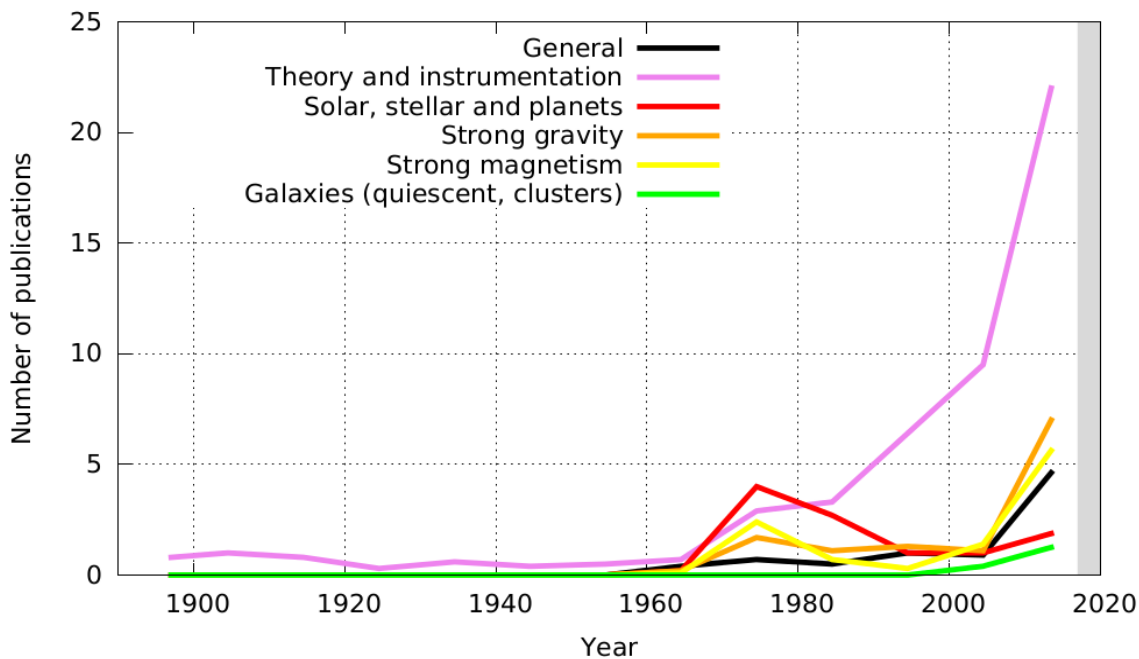

**Figure 2.** Number of publications focusing on astronomical X-ray polarimetry as a function of time. Several subcategories are highlighted: the black solid line represents general publications on X-ray polarimetry (reviews, historical notes); violet stands for publications concerning theory, instrumentation and satellites, red for papers on solar, stellar and exoplanet science; orange for papers on objects dominated by strong gravity effects (e.g., black holes); yellow for objects dominated by strong magnetic fields (e.g., neutron stars); and the green solid line stands for papers related to galaxies (e.g., galaxy clusters or the Milky Way).

We see in Figure 2 that a subcategory is clearly dominating the others between 1895 and 1968. The theoretical field was, as expected, predating any observation. Papers on specific celestial sources were almost inexistent as a turning point for the field of high energy astrophysics was the birth of X-ray astronomy with the first detection of a non solar source [2]. Moreover, at that time, it was still unclear if X-rays could be polarized and if this polarization could be detected. However, with the onset of the First Age of X-ray polarimetry (circa 1970 – 1990) and the first flights of rockets and satellites mounted with polarimeters, papers on specifics subjects started to dominate the theoretical and instrumental fields. Led by observations of solar flares and the Crab Nebula, the fields of "Solar, stellar and planets" and "Strong magnetism" science rose above the others. Numerical simulations also started with, e.g., the computation of the effect of special and general relativity onto the polarization of X-rays [21]. However, due to the lack of new observations, the interest of the community for X-ray polarization decreased in 1990. Only the field of "Theory and instrumentation" kept growing as, without new instruments or methods, it would be impossible to go on. Since the mid-80's, this field presents an almost exponential rise, culminating far above the other subcategories. It is only in the mid-2000's, after the development of the Gas Pixel Detector and Time Projection Chamber technologies, that the research community started to put effort on X-ray polarimetry again. Driven by the technological innovations, all the fields of research started to publish new predictions on what will be observed. We note that the "Solar, stellar and planets" and the "Galaxies" scientific fields are lagging behind. For the former case, this is due to the absence of envisioned mission dedicated to solar and stellar X-ray polarimetry in the next decade. A new mission would certainly uplift the community to work on this field again. In the later case, the very low X-ray fluxes, coupled to relatively low or inexistent polarization from galaxy clusters, naturally explains why the publication rate in this field is the lowest. Until the advent of extra-sensitive X-ray polarimeters, it is not mandatory to explore in greater details the expected polarization of quiescent galaxies (save the Milky Way, where important and feasible observations await us [22,23,24]).

## 3. Conclusions

In this paper, we saw that the number of publication related to astronomical X-ray polarimetry is growing at an important rate since the beginning of 2000. Driven by new technologies and mature polarimeters, all the scientific fields are publishing at an increasing rate and the growth of interest for high energy polarimetry is clearly measurable. The history of the field can be roughly summarized as follows: the First Age of (astronomical) X-ray polarimetry started in 1968 with the flight of the first Lithium-block Thomson-scattering polarimeter and ended circa 1980, with the termination of the OSO-8 mission. At that time, despite the fact that polarimeters sharing the focal plane of X-ray observatory missions suffered from the same difficulties that high-resolution spectroscopy did, the larger integration times needed by polarimeters to measure the polarized flux of cosmic sources and the mismatching between imaging/spectroscopic and polarimetric technologies were deemed too restrictive. This era was followed by a relatively quiescent period of 20 years, were technical innovations were slow yet steady. The advent of new technologies in the beginning of 2000 allowed the field to rebirth. This is the beginning of the Second Age for X-ray polarimetry. In comparison with the First Age, this era shows a slower development as it takes much more time to send a polarimeter in high altitudes or space than during the 70's. However, it also allows to build a stronger community of dedicated scientists. The Second Age is not yet at its pinnacle. From cliodynamical[1] reasoning, one would expect the publication rate to keep growing as we get closer to observational results. The succession of envisioned missions, that will cover at least 15 years, will maintain a high enthusiasm of the community for this scientific topic, ensuring a long and remarkable Second Age for X-ray polarimetry.

**Acknowledgments:** I am grateful to Martin C. Weisskopf and Enrico Costa who kindly agreed to share their historical knowledge on the subject with me, greatly improving the content of this paper. Additional information was kindly supplied by the two anonymous reviewers who participated to increase the quality of the manuscript. I would alike to acknowledge the Centre National d'Etudes Spatiales (CNES) for providing funding to achieve this paper through the postdoctoral grant "Probing the geometry and physics of active galactic nuclei with ultraviolet and X-ray polarized radiative transfer". I would also like to acknowledge financial support from the University of Strasbourg, from the Astronomical Observatory of Strasbourg, from the Programme National Cosmologie et Galaxies (PNCG) and from the Programme National Hautes Energies (PNHE).

---

1 Cliodynamics is a field of multidisciplinary research aimed at describing historical dynamics through mathematical models.

**Conflicts of Interest:** The authors declare no conflict of interest.